\documentclass[a4paper,fleqn,usenatbib,useAMS]{mnras}

\usepackage{mathptmx}

\usepackage[T1]{fontenc}
\usepackage{ae,aecompl}

\usepackage{graphicx}	
\usepackage{amsmath}	
\usepackage{amssymb}	

\title[Cooling X-ray halo of a rotating early-type galaxy]{Cooling in the X-ray halo of the rotating, massive early-type galaxy NGC~7049}

\author[A. Jur\'{a}\v{n}ov\'{a} et al.]{A. Jur\'{a}\v{n}ov\'{a}$^{1}$\thanks{E-mail: an.juranova@gmail.com},
	N. Werner$^{2,1,3}$,
	M. Gaspari$^{4}$\thanks{\it Einstein and Spitzer Fellow},
	K. Lakhchaura$^{2}$,
	P.E.J. Nulsen$^{5,6}$,
	\newauthor M. Sun$^{7}$,
	R.E.A. Canning$^{8}$\thanks{\it Einstein Fellow},
	S.W. Allen$^{8,9,10}$,
	A. Simionescu$^{11,12}$,
	J.B.R. Oonk$^{12,13}$,
	\newauthor T. Connor$^{14}$,
	M. Donahue$^{15}$
	\\
	$^{1}$Department of Theoretical Physics and Astrophysics, Masaryk University, Kotl\'a\v{r}sk\'a 2, 611 37 Brno, Czech Republic\\
	$^{2}$MTA-E\"otv\"os Lor\'and University Lend\"ulet Hot Universe Research Group, H-1117 P\'azm\'any P\'eter s\'et\'any 1/A, Budapest, Hungary\\
	$^{3}$School of Science, Hiroshima University, 1-3-1 Kagamiyama, Higashi-Hiroshima 739-8526, Japan\\
	$^{4}$Department of Astrophysical Sciences, Princeton University,  4 Ivy Lane, Princeton, NJ 08544-1001, USA\\
	$^{5}$Harvard-Smithsonian Center for Astrophysics, 60 Garden Street, Cambridge, MA 02138, USA\\
	$^{6}$ICRAR, University of Western Australia, 35 Stirling Hwy, Crawley,
	WA 6009, Australia \\
	$^{7}$Department of Physics and Astronomy, University of Alabama in Huntsville, Huntsville, AL 35899, USA\\
	$^{8}$Kavli Institute for Particle Astrophysics and Cosmology, Stanford University, 452 Lomita Mall, Stanford, CA 94305-4085, USA\\
	$^{9}$Department of Physics, Stanford University, 382 Via Pueblo Mall, Stanford, CA 94305-4060, USA\\
	$^{10}$SLAC National Accelerator Laboratory, 2575 Sand Hill Road, Menlo Park, CA 94025, USA\\
	$^{11}$SRON Netherlands Institute for Space Research, Sorbonnelaan 2, 3584 CA Utrecht, The Netherlands\\
	$^{12}$Leiden Observatory, Leiden University, PO Box 9513, 2300 RA Leiden, The Netherlands\\
	$^{13}$ASTRON, Netherlands Institute for Radio Astronomy, PO Box 2, NL-7990 AA Dwingeloo, the Netherlands\\
	$^{14}$The Observatories of the Carnegie Institution for Science, 813 Santa Barbara Street, Pasadena, CA 91101, USA\\
	$^{15}$Physics \& Astronomy Department, Michigan State University, East Lansing, MI 48824-2320, USA
}

\date{\today}

\pubyear{2018}

\begin{document}
	\label{firstpage}
	\pagerange{\pageref{firstpage}--\pageref{lastpage}}
	\maketitle
	
	\begin{abstract}
		The relative importance of the physical processes shaping the thermodynamics of the hot gas permeating rotating, massive early-type galaxies is expected to be different from that in non-rotating systems. Here, we report the results of the analysis of \textit{XMM-Newton} data for the massive, lenticular galaxy NGC~7049. The galaxy harbours a dusty disc of cool gas and is surrounded by an extended hot X-ray emitting gaseous atmosphere with unusually high central entropy. The hot gas in the plane of rotation of the cool dusty disc has a multi-temperature structure, consistent with ongoing cooling.
		We conclude that the rotational support of the hot gas is likely capable of altering the multiphase condensation regardless of the $t_{\rm cool}/t_{\rm ff}$ ratio, which is here relatively high, $\sim 40$.
		However, the measured ratio of cooling time and eddy turnover time around unity ($C$-ratio $\approx 1$) implies significant condensation, and at the same time, the constrained ratio of rotational velocity and the velocity dispersion (turbulent Taylor number) ${\rm Ta_t} > 1$ indicates that the condensing gas should follow non-radial orbits forming a disc instead of filaments. This is in agreement with hydrodynamical simulations of massive rotating galaxies predicting a similarly extended multiphase disc. 
	\end{abstract}
	
	
	
	
	\section{Introduction}
	
	The long-lasting presence of hot gas in early-type galaxies, its connection to the cold/cool phase and its role in galaxy evolution are still not fully understood. Observations of the most massive ellipticals suggest that the cold interstellar medium in these systems is produced mainly via cooling from the hot X-ray emitting atmospheres \citep{Werner2014,lakhchaura2018}. Conditions required for the development of thermal instabilities in the hot phase are likely different when rotational support prevents the gas from moving in radial directions and thus alters the cooling flow.
	
	The hot X-ray emitting gas in fast-rotating galaxies has a systemically lower surface brightness and mean temperature than the hot gas in the non-rotating systems of the same mass \citep[e.g.][]{Negri2014}. This is most likely the result of the combined effect of the centrifugal barrier in the rotating atmosphere and the decreased depths of the effective gravitational potentials due to the rotational support. Apart from affecting the X-ray luminosities and shapes of the hot atmospheres \citep{Brighenti1996,Brighenti1997, Hanlan2000, Machacek2010}, rotation should also influence the conditions that govern thermal instabilities in the hot gas. \citet{Gaspari2015} showed that the top-down multiphase condensation process -- also known as chaotic cold accretion (CCA) -- changes in rotation-dominated atmospheres.  When the ratio of the rotational velocity and the velocity dispersion of the hot gas (also called turbulent Taylor number) ${\rm Ta_t} \equiv v_{\rm rot}/\sigma_{\rm v} > 1$, the condensation will produce an extended multiphase disc instead of thin filaments, while suppressing the accretion rate onto the central supermassive black hole (SMBH) due to the centrifugal barrier \citep{Gaspari2017}. 
	Moreover, the criterion $ C \equiv t_{\rm cool}/t_{\rm eddy} \approx 1 $ is expected to be related to the extent of the condensation region \citep{Gaspari2018}. To study thermally unstable cooling in rotating systems, we observed NGC~7049, a rotating early-type galaxy with an extended hot X-ray emitting atmosphere that also harbours a disc of cold/warm gas, which might have formed as a result of cooling from the hot phase onto non-radial orbits. 
	
\begin{figure}
		\centering
		\includegraphics[width=0.7\linewidth]{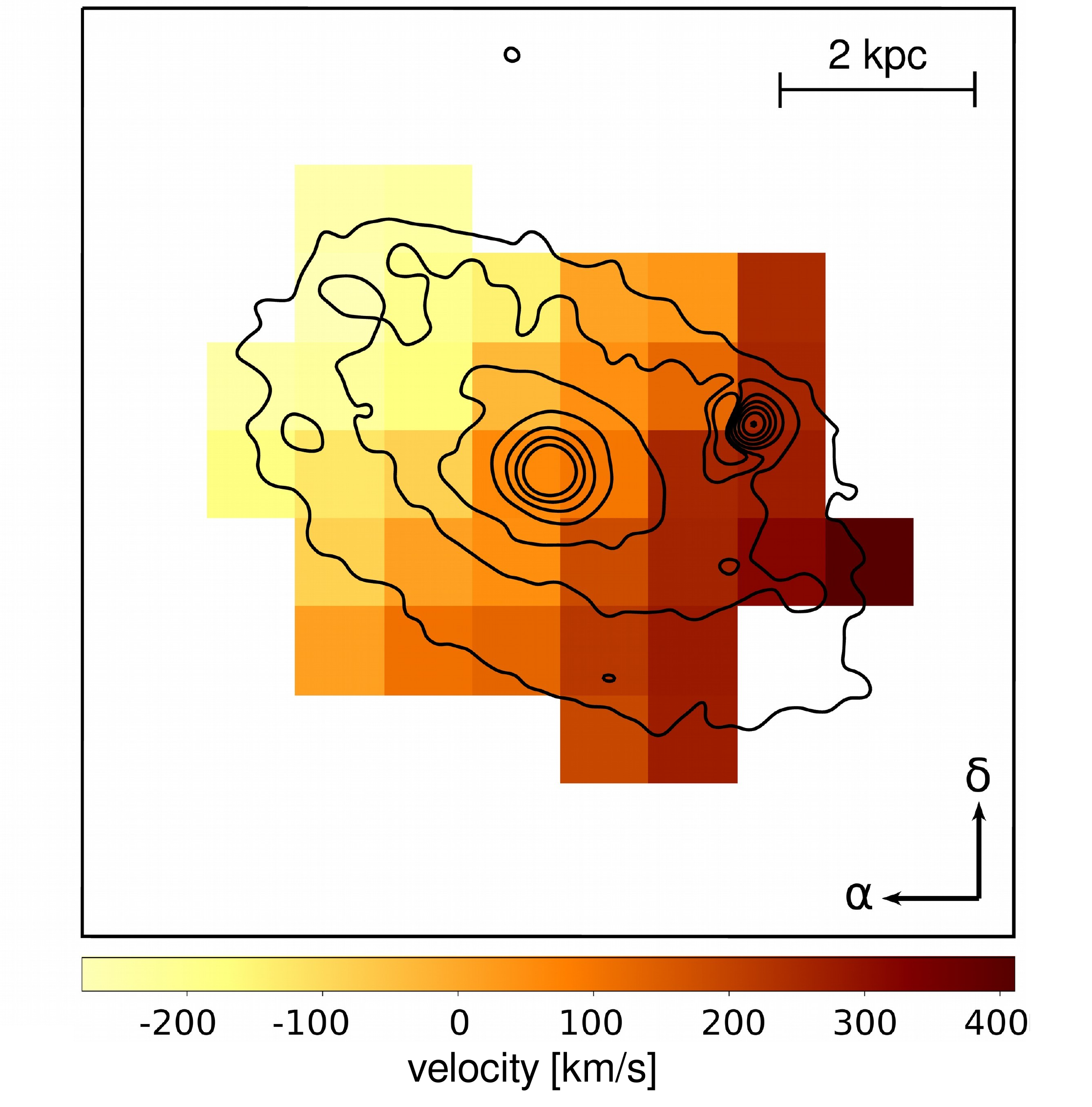}
		\caption{$\mathrm{[C\,\textsc{ii}]} $ line-of-sight velocity map of NGC~7049 based on \citet{Werner2014}. Values are relative to the systemic velocity of the galaxy. The velocity distribution of the $\mathrm{[C\,\textsc{ii}]} $ emitting gas indicates the presence of a disc, rotating with a velocity of $v_{\rm rot}\sim200$~km~s$^{-1}$.}
		\label{fig:cii_v}
	\end{figure}

	Here, we present an analysis of \textit{XMM-Newton} data for the massive fast rotating unbarred lenticular (SA$0^0$) galaxy NGC~7049, the brightest member of a small group of five to six galaxies. The total mass $M_{\rm p}$ (\citealt{Heisler1985}, equation 11) of the group is $ \log M_{\rm p}/M_{\odot} = 13.16 $ \citep{Makarov2011}. The target was chosen based on the sample of \citet{Werner2014} as an X-ray bright massive galaxy with extended X-ray emission and significant angular momentum. It is inclined by $ \sim 30^{\circ} $ with respect to the line of sight and is relatively nearby \citep[$29.9~\mathrm{Mpc}$;][]{Tonry2001}. \citet{Werner2014} present measurements of optical $ \mathrm{H}\,\alpha \mathrm{+[N\,\textsc{ii}]} $ emission, a tracer of warm ionised gas, observed with the Southern Astrophysical Research (SOAR) telescope, and far-infrared $\mathrm{[C\,\textsc{ii}]\lambda157~\umu m} $, $\mathrm{[O\,\textsc{i}]\lambda 63~\umu m} $, and $\mathrm{[O\,\textsc{i}b]\lambda 145~\umu m} $ emission of cold ($ \sim100~\mathrm{K} $) atomic gas measured with the \textit{Herschel} Photodetector Array Camera \& Spectrometer (PACS). Both the optical and far-infrared emission components have disc-like morphology and extend out to $ r \sim 3.5~\mathrm{kpc} $. The velocity distribution calculated from the $\mathrm{[C\,\textsc{ii}]} $ line indicates that the cold gas rotates with a velocity of $ v_{\rm rot, [C\,\textsc{ii}]} \approx 200~\mathrm{km\, s^{-1}}$ (see Fig. \ref{fig:cii_v}).
	The radio emission with $ L_{\mathrm{radio}} = 8.4\times 10^{37}~\mathrm{erg~s^{-1}} $ suggests the presence of radio-mode activity of the central active galactic nucleus (AGN). The  galaxy has low star formation rate  of $ 0.177~\mathrm{M_{\odot}~yr^{-1}}$ \citep{Carlqvist2013}, effective radius of $ 4.4~\rm kpc $ \citep{Blakeslee2001}, absolute magnitude of the bulge in K-band is $ -24.81~\rm mag $ and its K-band bulge-to-total flux ratio is $0.81$; the latter two are presented in \citet{Laurikainen2010}.
	
	At the distance of $29.9~\mathrm{Mpc}$  \citep[][]{Tonry2001}, the angular scale is $6.89$ arcseconds per kpc. Throughout the analysis, we use the Solar abundances of \citet{Lodders2009}. All results in the following sections are presented with $ 1\sigma $ error bars. In the data analysis, we also used Python \citep{Python} and its specialised libraries Scipy \citep{Scipy}, Numpy \citep{Numpy} and Matplotlib \citep{Matplotlib}.
	
	The paper is structured as follows. In Sect. \ref{sec:analysis} we describe analysis of the \textit{XMM-Newton} data, in Sect. \ref{sec:results} we show the main results, focusing on X-ray gas morphology (Sect. \ref{sec:morphology}) and global properties of the gas (\ref{sec:global}), we then proceed to their inspection in radial profiles (\ref{sec:radial}) and further on dissected into sectors corresponding with the plane of rotation and the rotational axis (\ref{sec:sectors}). In Sect. \ref{sec:cooling} we present radial profiles of the main criteria which are expected to predict the thermodynamic and kinematic state of the X-ray emitting halo. In Sect. \ref{sec:discussion} we discuss our findings and present our conclusions in Sect. \ref{sec:conclusions}.

	\section{Observation and data analysis}\label{sec:analysis}
	
	The $71~\rm ks$ \textit{XMM-Newton} observation of NGC~7049 was performed on April 3 2015 (dataset ID: 0743930101). We processed the raw data with the Science Analysis System (SAS) version 15.0.0. To filter out the time periods affected by soft-proton flares, we excluded the data where the count rate deviated  from the mean by more than $1.5~\sigma$. The observation is strongly affected by flares, with the net exposure time reduced to less than 50 per cent of the observation (see Table \ref{tab:times}).
	
	\begin{table}
		\centering
		\caption{The total observing time $ t_{\rm tot} $ and the net time after soft-proton filtering $ t_{\rm net} $ for all instruments used.}
		\label{tab:times}
		\begin{tabular}{lccr}
			\hline
			& MOS1 & MOS2 & pn\\
			\hline
			$ t_{\rm tot}~[\rm ks] $ & 69.7 & 69.7 & 65.7\\
			$ t_{\rm net}~[\rm ks] $ & 27.2 & 32.3 & 18.0\\
			\hline
		\end{tabular}
	\end{table}
	
	The selection of regions for spectral extraction and point source removal was done based on visual inspection, taking into account the instrumental point-spread function (PSF). The position of the AGN ($ \alpha = 21^{\mathrm{h}}~19^{\mathrm{m}}~0.17^{\mathrm{s}},\ \delta = -48^\circ~33\arcmin~43\farcs 45 $) was determined using a short, $2.2~\mathrm{ks}$ (Obs ID: 5895), observation with the \textit{Chandra X-ray Observatory}, which has a superior spatial resolution. For each point source identified by {\it XMM-Newton} or {\it Chandra}, we excluded the data within a circular region with a radius of  $ 15~\mathrm{arcsec} $. We studied the properties of the hot gas emission in six concentric annuli with the largest reaching out to $ r=175~\mathrm{arcsec} $ (approximately $ 25.4~\mathrm{kpc} $). An additional outer annulus extending out to $r=237.5~\mathrm{arcsec}$ ($ 34.4~\rm{kpc} $) was used to account for the emission from the outskirts of the system. These annuli were then divided into four quadrants, for which the orientation was chosen to match the axes of the best-fit ellipse to the optical emission of the galaxy. 
	
	Given the relatively large widths of our extraction regions, the PSF of \textit{XMM-Newton} does not affect the derived profiles significantly. With most of the observed emission being soft, below $1.2~\mathrm{keV}$ as implied below, the energy-dependence of the PSF is also negligible.
	
	The spectral fitting was performed in the  $ 0.3 - 5.0~\mathrm{keV} $ energy range. While the soft X-ray band is dominated by the emission of the gaseous atmosphere of NGC~7049, the emission at higher energies is dominated by the power-law-like emission component of the unresolved population of low-mass X-ray binaries (LMXBs), CV stars and the cosmic X-ray background. The data in the energy range of $ 1.38 - 1.60~\mathrm{keV} $ were ignored due to possible contamination with instrumental line emission. The data were binned to at least one count per bin and the fitting was performed using C-statistic \citep{Cash1979}.
	
	Part of the spectral analysis was performed using the SPEX v. 3.04 spectral fitting package \citep{kaastra1996} with SPEXACT v. 3. To account for Galactic absorption, we used the \texttt{hot} model with a particle column density of $ N_{\rm H} = 2.70\times10^{20}~\rm cm^{-3} $ provided by the Leiden/Argentine/Bonn Survey \citep{LAB}. The emission of the hot gas was modelled with a collisional ionisation equilibrium plasma model (\texttt{cie}), assuming a redshift $ z=0.0073 $. The \texttt{cie} model in SPEX can also be used as a~differential emission measure model, where the emission measure $ Y $ as a~function of $ k_{\mathrm{B}}T $ has a~Gaussian shape (GDEM). The GDEM model has an additional free parameter $\sigma_{T}$, which is the width of the Gaussian expressed in keV \citep[for a more detailed description of the GDEM model, see][]{deplaa2017}. 
	
	The deprojection analysis was performed with the XSPEC spectral fitting package  \citep[][v. 12.9.1 with atomic database AtomDB 3.0.7]{Xspec} using both direct spectral deprojection \citep[DSDEPROJ,][]{DSDEPROJ} and the model \texttt{projct} for comparison. To model the hot gas here, we used the model \texttt{vapec}. The X-ray emission of the population of unresolved stellar sources in NGC~7049 is modelled by a power-law emission model with a photon index $ \Gamma = 1.6 $ \citep[e.g. see][]{Irwin2003}, with the normalisation left as a~free parameter. For the direct spectral deprojection, we had to use $\chi^2$ statistics, which is required for the implementation of the DSDEPROJ model, and we binned the spectra to 25 counts per bin. To account for the projected emission from beyond the last annulus, we scaled down the normalisation of the outermost spectrum by extrapolating the particle number density based on a $\beta$-model

	\begin{equation}\label{eq:beta}
	n(r) = n(0) \left[ 1+ \left(\frac{r}{r_{\mathrm{c}}}\right)^2\right]^{-3\beta/2}
	\end{equation}
	
	\noindent \citep{Cavaliere1978}, where $ r_{\mathrm{c}} $ is the core radius, which we fitted to the data points obtained from the deprojected spectra, leaving out the last one.

	\section{Results}\label{sec:results}
	
	\subsection{X-ray morphology}\label{sec:morphology}
\begin{figure*}
		\centering
		\includegraphics[width=0.9\linewidth]{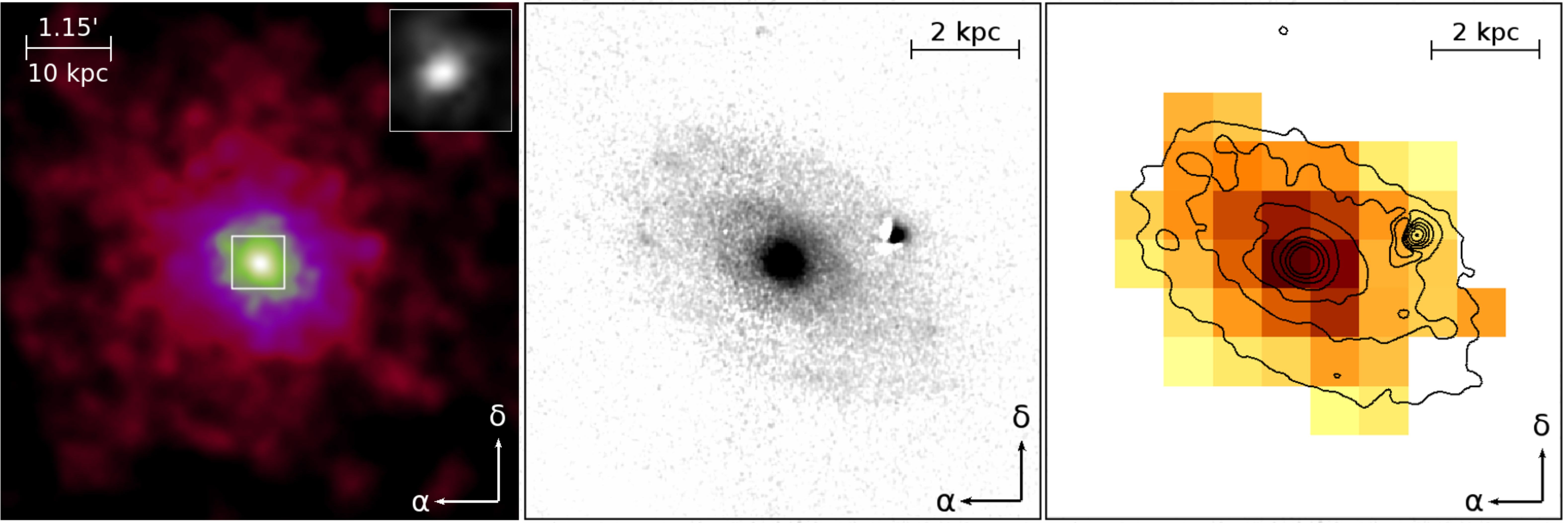}
		\caption{Left: \textit{XMM-Newton} X-ray ($ 0.3 - 2.0~\rm{keV} $) image of NGC~7049 composed of data from all EPIC instruments, exposure-corrected and adaptively smoothed. The black-and white image in the upper-right corner is a detail of the central region observed by the \textit{Chandra X-ray Observatory} in a 5\,ks exposure. Centre: emission of warm ionised gas in $ \mathrm{H}\,\alpha \mathrm{+[N\,\textsc{ii}]} $ lines observed in NGC~7049 by the SOAR telescope. Right: $\mathrm{[C\,\textsc{ii}]}$ line flux obtained by {\it Herschel} PACS with contours of the $ \mathrm{H}\,\alpha \mathrm{+[N\,\textsc{ii}]}$ emission overlaid.}
		\label{fig:xray}
	\end{figure*}

	The X-ray image of NGC~7049 (Figure \ref{fig:xray}) reveals that the shape of the X-ray emission surrounding the galaxy deviates from circular symmetry only slightly. We determined its projected ellipticity, $ \epsilon_{\rm X} $, using the CIAO (version 4.9, \cite{CIAO}) fitting tool Sherpa and the \texttt{beta2d} model, yielding $ \epsilon_{\rm X} = (0.126\pm0.004) $. The projected ellipticity of the stellar component, $ \epsilon_{\star} $, determined in the DSS IIIaJ band ($ \sim 468~\rm nm $) is more pronounced: $ \epsilon_{\star} = (0.25 \pm 0.02)$. The major axes of the two emission components lie in the same direction. 
	
\begin{figure}
		\centering
		\includegraphics[width=1\linewidth]{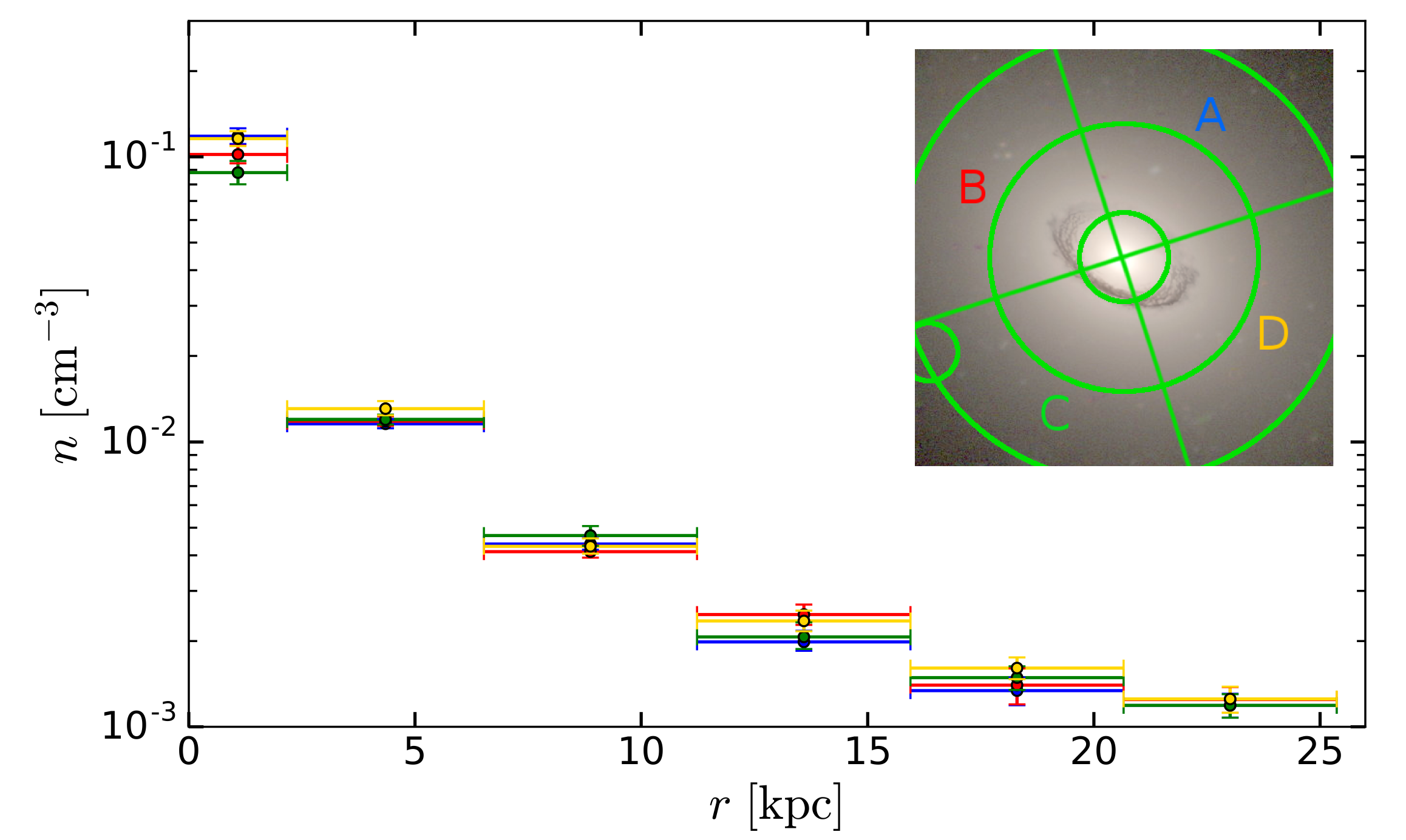}
		\caption{Projected particle densities measured in four directions: red and yellow are approximately parallel with, and blue and green are perpendicular to the plane of rotation of the cold gas disc. For clarity, the disc and the orientation of sectors are shown in a small HST image in the upper right corner.} \label{fig:sfb}
	\end{figure}

	The normalisation of the \texttt{cie} model in SPEX is equal to the emission measure of the gas $Y= n_{\mathrm{H}}n_{\mathrm{e}}V $. Having this value for four quadrants at six distances from the centre determined using spectral fits, we were able to constrain the projected particle density of the plasma. The results, calculated assuming a constant line-of-sight column depth of $20~\mathrm{kpc}$, are displayed in Figure \ref{fig:sfb} and show that there is no significant departure from radial/spherical symmetry. Because the LMXB component follows the stellar distribution, the ellipticity seen in the X-ray image might be, at least in part, due to the X-ray emission of unresolved stellar sources.
	
	\subsection{Global features}\label{sec:global}
	First, we derived the basic properties of the galaxy using its global spectrum extracted from an annular region spanning $r= 15\arcsec - 165\arcsec $. The innermost region was excluded due to the contamination by the X-ray bright AGN, leaving approximately 8260 counts detected within the extraction area after background subtraction. Because the isothermal CIE model does not provide a good fit to the global spectrum, we derived the results listed below using the GDEM model, which only adds one free parameter to our fit. The emission-weighted temperature of the gas is $ k_{\mathrm{B}}T = 0.43^{+0.02}_{-0.01}~\mathrm{keV}$ with $ \sigma_{T} = 0.21\pm 0.03~\mathrm{keV}$. The relatively high signal-to-noise ratio of the data allowed us to constrain the overall emission-weighted metallicity of the galactic atmosphere to $ Z = 0.7^{+0.2}_{-0.1}~ Z_{\odot}$.
	
	\subsection{Radial profiles}\label{sec:radial}
	
	The central region is contaminated both by a power-law-like emission from unresolved stellar sources in the galaxy and by a power-law-like AGN emission. Because the data do not allow us to constrain the spectral indices of both power-law components independently, we fitted their emission with a single power-law with a best-fit photon index $ \Gamma_{\mathrm{AGN+LMXB}} = 1.7\pm 0.1 $ (fixed at this value for the deprojection analysis). From the best-fit parameters of the power-law component, we constrain the X-ray luminosity of the AGN to be $ L_{\rm X, AGN}\lesssim 6.2 \times 10^{39}~\mathrm{erg~s^{-1}} $. Assuming the Magorrian relation \citep{Kormendy2013}, the SMBH mass $\approx 10^9~\mathrm{M}_{\odot}$, which implies a highly sub-Eddington rate of $\approx 10^{-7}$. 
	
	\begin{figure}
		\centering
		\includegraphics[width=1.\linewidth]{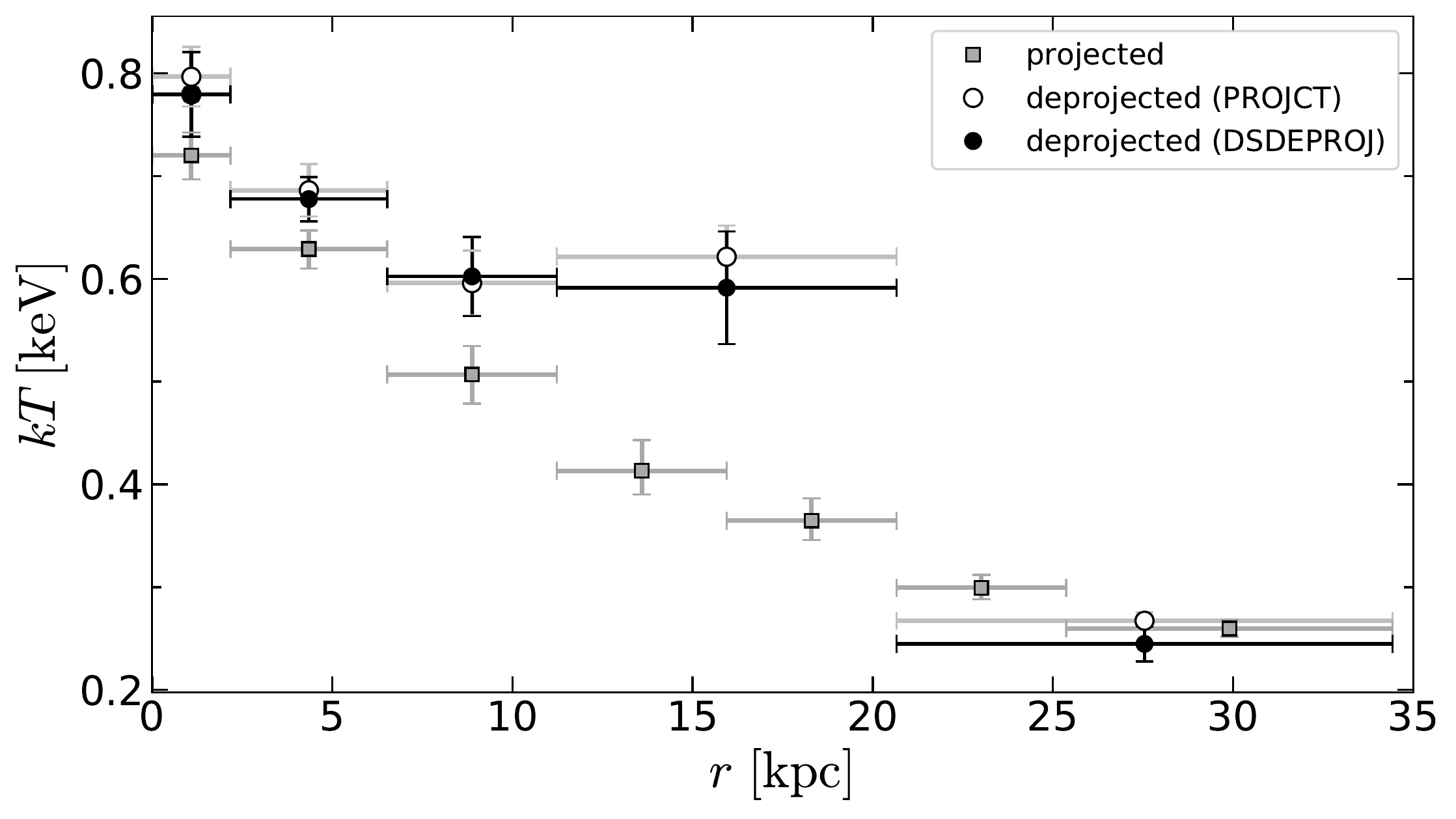}
		\caption{The radial profile of the hot gas temperature in NGC~7049 derived from deprojected (black circles for deprojection with DSDEPROJ and white circles for model \texttt{projct}) and projected spectra (grey squares). To increase the number of counts and mitigate the ringing effect in the deprojection analysis, the spectra from annuli outside the third one were tied by two and thus are displayed as only two points for each deprojection method.}
		\label{fig:kttext}
	\end{figure}
	
	A common systematic uncertainty when fitting spectra of low temperature systems is the anti-correlation of metallicity and normalisation, which often leads to gross underestimates of the metallicity \citep[see e.g.][]{buote2000,werner2008}. Therefore, based on the global fit, we fixed the metallicity to $ 0.7~Z_{\odot} $ and assumed it to be constant as a function of both radius and azimuth. 
	
	In Figure \ref{fig:kttext}, we present the temperature profile where the results obtained from the deprojection analysis are shown as circles (white for deprojection with \texttt{projct} and black corresponding to results from DSDEPROJ) and, for comparison, the values determined from the projected spectra are indicated as grey squares. 
	For deprojection, taking into account lower number of counts resulting from the deprojection and the use of $\chi^2$ statistics in case of DSDEPROJ, we combined the spectra of $4^{\rm th}$ and $5^{\rm th}$, and $6^{\rm th}$ and $7^{\rm th}$ radial bins. Unless otherwise stated, further on we use the DSDEPROJ deprojection method.
	
	\begin{figure}
		\centering
		\includegraphics[width=1.\linewidth]{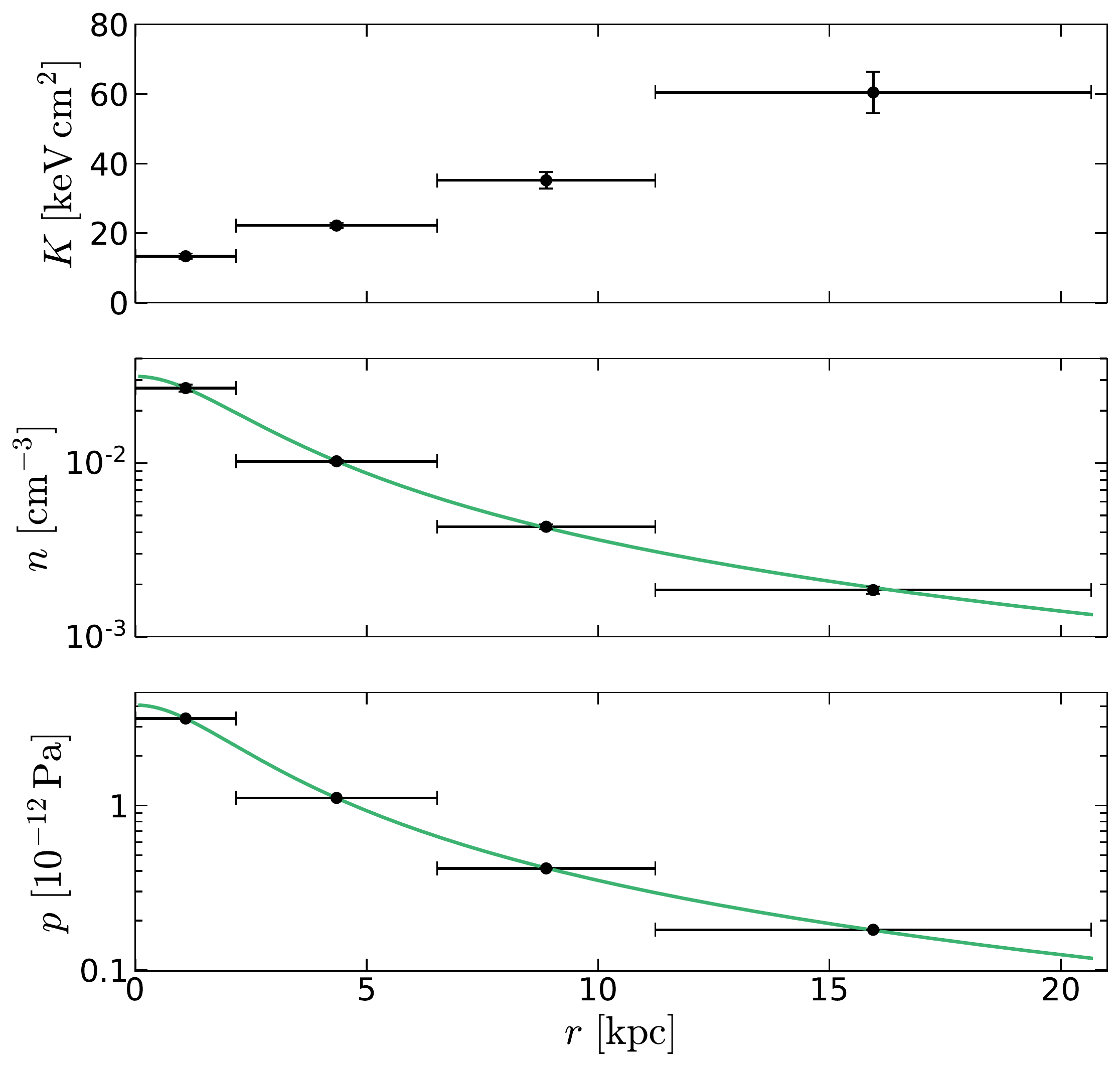}
		\caption{Profile of deprojected entropy $ K = k_{\mathrm{B}}Tn_{\mathrm{e}}^{-2/3} $, particle density $ n $ and pressure $ p = nk_{\mathrm{B}}T $. The solid green curves are the best-fit $ \beta $-models (see eq. \ref{eq:beta} and eq. \ref{eq:beta_p}). We note that the errors, especially in the bottom plot, are too small to be visible.}
		\label{fig:Knptext}
	\end{figure}
	
	From the best-fit normalisations and temperatures obtained from the deprojected spectra we calculated the particle number densities, entropies, and pressures. We define the entropy as $ K = k_{\mathrm{B}}Tn_{\mathrm{e}}^{-2/3} $, where $ n_{\mathrm{e}} $ is the electron number density and $ k_{\rm B} $ is the Boltzmann constant. The pressure is $ p = nk_{\mathrm{B}}T $, where $ n $ is the total particle number density $ n = 1.92\,n_{\mathrm{e}} $. Because the value of the spectral normalisation determined for the last annulus depends critically on our assumptions about the emission beyond its outer boundary, it has significant systematic uncertainties. Therefore, to remain conservative, we do not use the normalisation from the last bin in the rest of the paper. The best-fit profiles are shown in Figure \ref{fig:Knptext}.
	
	We fitted the deprojected particle density profile with a single $ \beta $-model (\ref{eq:beta}). We left $ \beta $, $ n(0) $ and $ r_{\mathrm{c}} $ as free parameters yielding $ \beta = 0.47 \pm 0.01 $ and $ n(0) = (0.031 \pm 0.001)~\rm cm^{-3}$. The model is plotted along with the data in Figure \ref{fig:Knptext}.\\
	
	The pressure profile was modelled with a similar $ \beta $-model:
	
	\begin{equation}\label{eq:beta_p}
	p(r) = p(0) \left[ 1+ \left(\frac{r}{r'_{\mathrm{c}}}\right)^2\right]^{-3\beta'/2}.
	\end{equation}
	
	The best-fit parameters of the fit are: $ p(0) = (4.07 \pm 0.04)\times 10^{-12}~\mathrm{Pa} $, $r'_{\rm c} = (2.06 \pm 0.03)~\mathrm{kpc}$, $\beta' = (0.51 \pm 0.01)$.
	
	\subsection{Profiles in sectors}\label{sec:sectors}
\begin{figure}
		\centering
		\includegraphics[width=1.\linewidth]{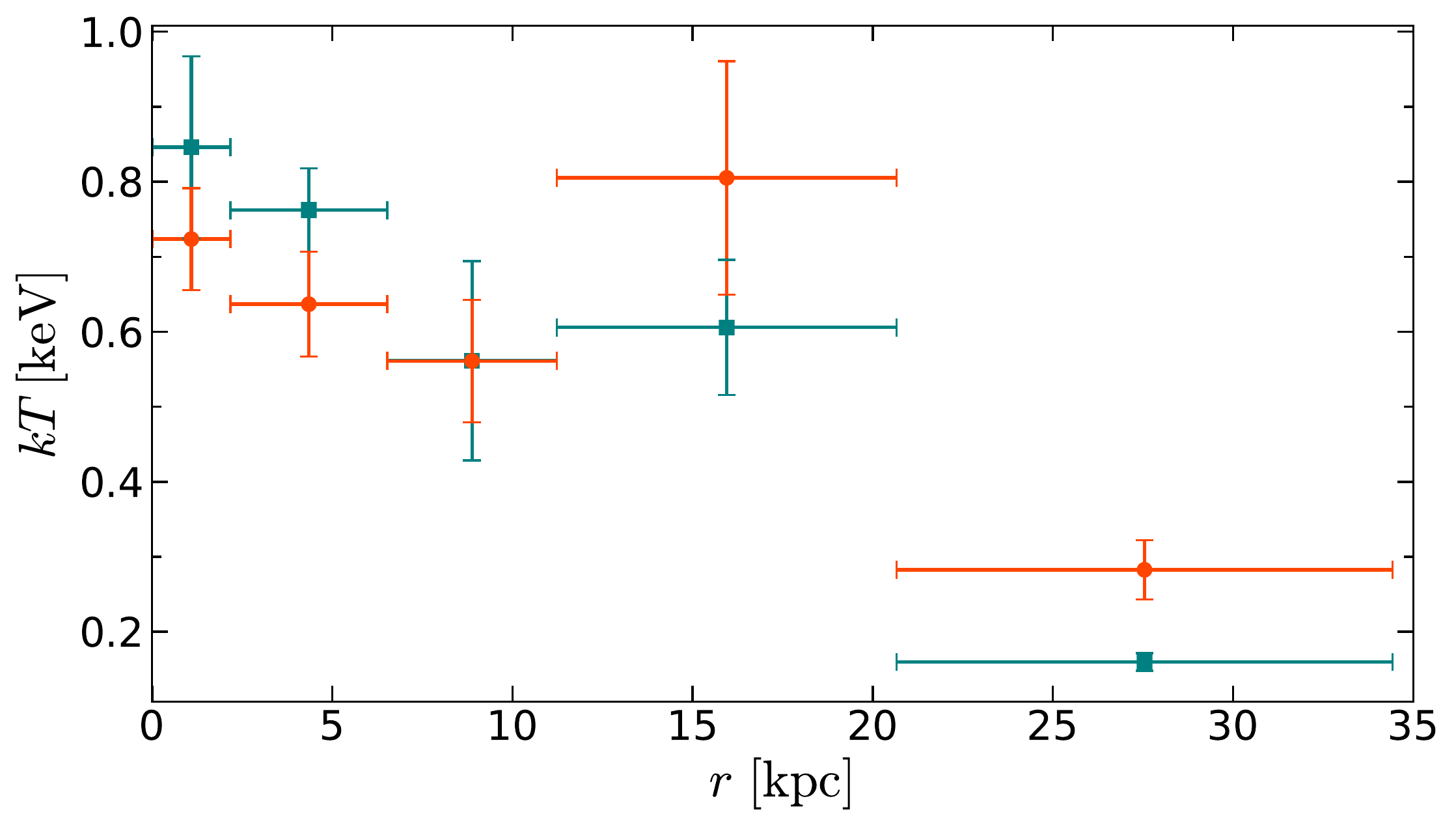}
		\caption{Deprojected temperature profiles determined in sectors along the rotational axis (B+D) and in a perpendicular direction (A+C) shown as turquoise squares and orange dots, respectively.}
		\label{fig:kt_sectext}
	\end{figure}

\begin{figure}
		\centering
		\includegraphics[width=1.\linewidth]{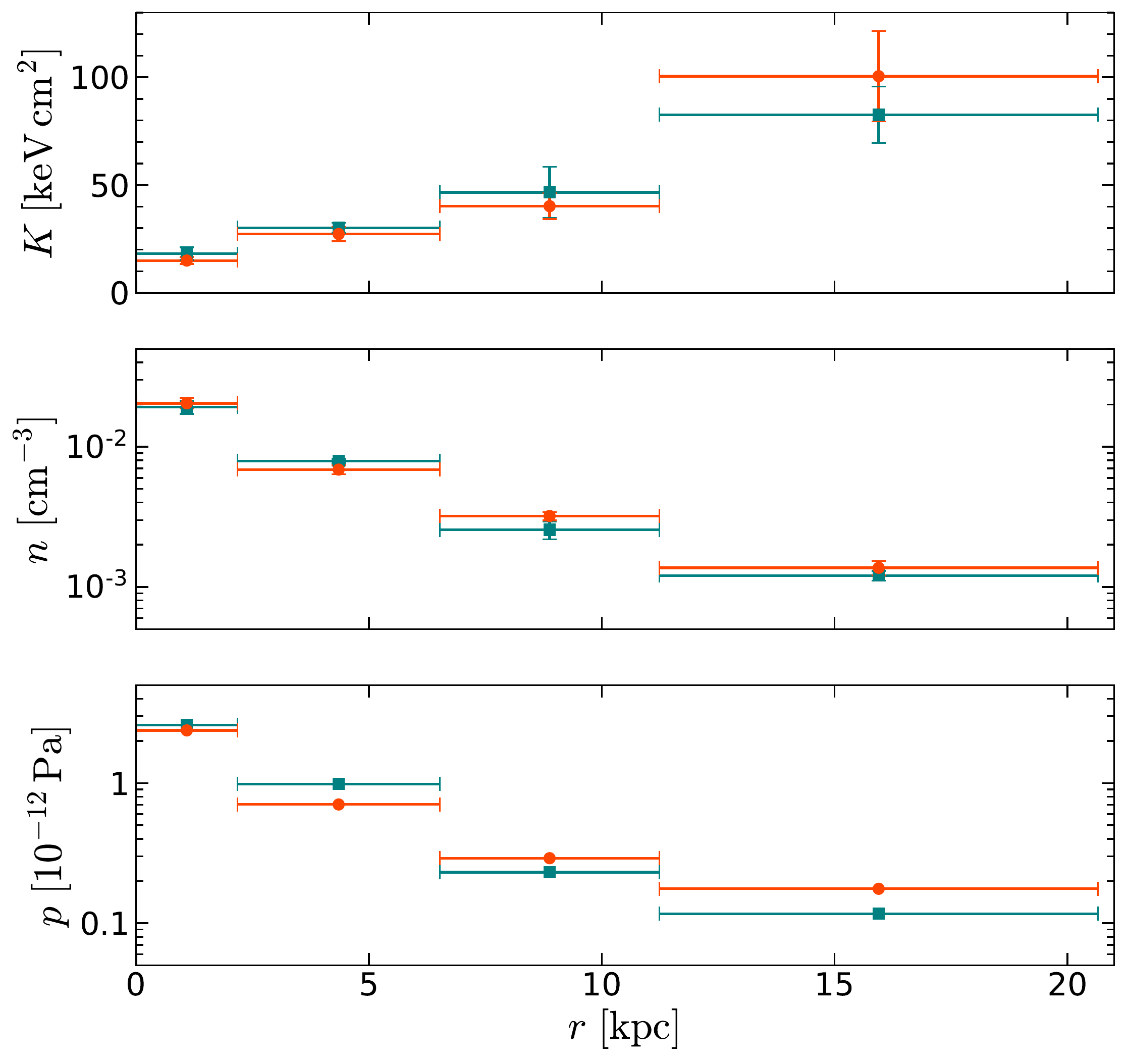}
		\caption{Deprojected profile of entropy $ K = k_{\mathrm{B}}Tn_{\mathrm{e}}^{-2/3} $, particle density $ n $ and pressure $ p = nk_{\mathrm{B}}T $ in sectors along the rotational axis (turquoise squares) and the perpendicular direction (orange dots). We note that errors especially in the bottom plot are too small to be visible.}
		\label{fig:Knp_sectext}
	\end{figure}

	Subdividing the emission into sectors allows us to create profiles for the hot gas in the plane of rotation (named B and D in accordance with the notation in Figure \ref{fig:sfb}) and in the direction perpendicular to it (A and C). We remind that the inclination of the plane of rotation is $ i\approx 30^{\circ} $. The results from the deprojection are displayed in Figures \ref{fig:kt_sectext} and \ref{fig:Knp_sectext}, where the data for sectors A+C are plotted using turquoise squares, while sectors B+D are indicated with orange dots. As in the case of the azimuthally averaged radial profiles, for the outermost bin we only show the best-fit value of the temperature, which is presumably less affected by our assumptions about the gas distribution further out.
	
	These results indicate that the gas distribution is consistent with spherical symmetry.

	\subsection{Cooling process} \label{sec:cooling}
	The rotating disc of cool atomic gas that spans a few kiloparsecs may indicate an ongoing cooling of the X-ray gas onto non-radial orbits in the plane of rotation. To test whether such a process is present and detectable by means of X-ray spectroscopy, we searched for multi-temperature gas both along the plane of the disc and in the perpendicular direction. 
	
	As the gas emission is proportional to the density squared, it is expected to form cooling clumps of different temperatures, giving rise to a complex spectrum. Such spectrum can then be described by a~differential emission measure distribution, such as the GDEM model in the SPEX spectral fitting package.  
	
	We measured a non-zero $ \sigma_{T} $ with a higher than $ 99.73~\% $ significance in the B+D sector of the second annulus ($ 15\arcsec - 45\arcsec$, which corresponds to $ 4.3 - 8.7~\mathrm{kpc} $, denoted as $ \mathrm{(B+D)}_{\mathrm{(2)}} $ further on), whilst in the perpendicular direction, $ \mathrm{(A+C)}_{\mathrm{(2)}} $, no such feature was observed. More precisely, we measured $ \sigma_{T,\mathrm{B+D}} = 0.21^{+0.05}_{-0.06}~\rm keV $ with a mean temperature $ k_{\mathrm{B}}T_{\mathrm{B+D}} = 0.51^{+0.03}_{-0.03}~\mathrm{keV}$ in $\mathrm{(B+D)}_{\mathrm{(2)}} $ and $ \sigma_{T,\mathrm{A+C}} = 0.00^{+0.07}_{-0.00}~\rm keV $ in region $ \mathrm{(A+C)}_{\mathrm{(2)}} $ in projected spectra. Implicitly, the range of temperatures in the plasma of $\mathrm{(B+D)}_{\mathrm{(2)}} $ is approximately $ (0.30-0.72)~\mathrm{keV}$.
	
	We obtained qualitatively similar results with a two-temperature model and with a model of a power-law-like differential emission measure distribution.
	Due to the low quality of the data, we cannot discriminate between different multi-temperature models.
	
	To place constraints on the cooling rate, we fitted an isobaric cooling flow model (assuming the plasma is not supported by a~magnetic field) described by the emission measure as
	
	\begin{equation}\label{eq:cflow}
	\frac{\mathrm{d}Y}{\mathrm{d}T} = \frac{5}{2}\frac{\dot{M}k_{\rm B}}{\mu m_{\mathrm{H}}{\Lambda}(T)}
	\end{equation}
	
	\noindent to the spectrum of $ \mathrm{(B+D)}_{\mathrm{(2)}} $ resulting in mass deposition rate $ \dot{M} \sim 10^{-1}~\mathrm{M}_{\odot}~\mathrm{yr}^{-1}  $. More accurate estimate could not be made since the cooling function $ \Lambda $ \citep[with values tabulated in][]{Schure2009} in eq. (\ref{eq:cflow}) is strongly dependent on the abundances of heavier elements, which are highly uncertain in this case.
	The mass deposition rate was estimated using the metallicity $ Z = 0.7~Z_{\odot} $, measured from the global spectrum, in the temperature range of $ 0.1 - 0.9~\mathrm{keV} $.
	As a reference, the value of the classical cooling rate (upper limit to the actual cooling rate) is $ \dot{M} = 2  m_{\rm H} \mu L_{\rm X}/(5k_{\rm B} T) \approx 0.5~\mathrm{M}_{\odot}~\mathrm{yr}^{-1} $, where we used the X-ray luminosity $L_{\rm X} = 5.51\times10^{40}~\mathrm{erg~s^{-1}}$ and $ k_{\rm B} T = 0.43~\rm{keV}$.
	
	\begin{figure}
		\centering
		\includegraphics[width=1.\linewidth]{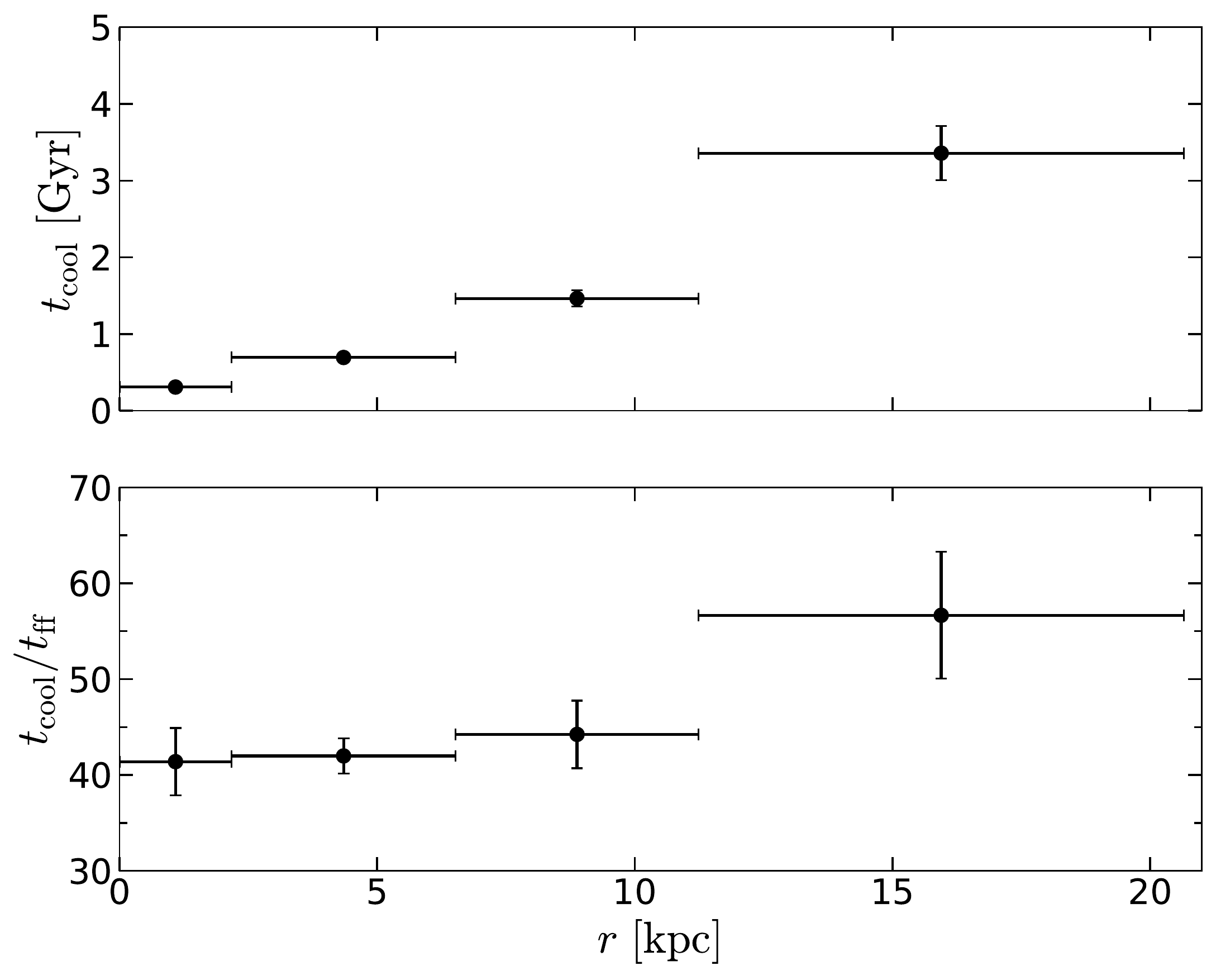}
		\caption{Profiles of cooling time and cooling time to free-fall time ratio evaluated for the four distances from central AGN from deprojected spectra. According to \citet{Sharma2012}, when the ratio $ t_{\mathrm{cool}}/t_{\mathrm{ff}} $ falls below $10$, thermal instabilities can form in the X-ray gas.}
		\label{fig:tc_tff_text}
	\end{figure}
	
	\subsubsection{$t_{\rm cool}/t_{\rm ff}$}
	Recent simulations \citep[e.g.][]{McCourt2012,Sharma2012} suggest that the classical thermal instability (TI) grows non-linear in the hot medium if the ratio of the cooling time to the free-fall time falls below $t_{\mathrm{cool}}/t_{\mathrm{ff}}\sim10$.
	
	We calculated the free-fall time simply as $t_{\mathrm{ff}} = \sqrt{2r/g} $, where the gravitational acceleration $ g $ is derived from the functional form of the pressure profile as 
	
	\begin{equation}
	g = -\frac{1}{\rho} \frac{\mathrm{d}p}{\mathrm{d}r} = -\frac{1}{nm_{\mathrm{H}}\mu} \frac{\mathrm{d}p}{\mathrm{d}r},
	\end{equation}
	
	\noindent  with mean atomic weight $ \mu=0.62 $. The cooling time, defined as
	
	\begin{equation}\label{eq:tcool}
	t_{\mathrm{cool}} = \frac{\frac{3}{2}(n_{\mathrm{e}} + n_{\mathrm{i}})k_{\mathrm{B}}T}{n_{\mathrm{e}} n_{\mathrm{i}} {\Lambda}(T)},
	\end{equation} 
	
	\noindent was calculated with the cooling function, $\Lambda(T)$, of \citep{Schure2009} for a metallicity of  $0.7~Z_{\odot} $ and the ion density $n_{\mathrm{i}}=0.92n_{\mathrm{e}}$. The results for $ t_{\mathrm{cool}} $ and the ratio $ t_{\mathrm{cool}}/t_{\mathrm{ff}} $ are displayed in Figure \ref{fig:tc_tff_text}. Although multiphase gas -- including $\mathrm{H}\,\alpha  $ and $[\mathrm{N\,\textsc{ii}}]$ emission -- is clearly observed in NGC~7049, the criterion for the development of cooling instabilities does not hold in this system. 
	
	On the other hand, other studies \citep[e.g.][]{Gaspari2012} show that the TI-ratio threshold is not a demarcation line, but has a large scatter, in some instances condensation can be seen with ratios up to $50$. Another point to consider is the time hysteresis: as warm gas quickly drops out of the hot plasma, the entropy of the more diffuse and hotter plasma phase rapidly increases, thus inducing larger TI-ratio. In other words, the current TI-ratio could be a predictor of future condensation, but not necessarily of the previous phase which generated the currently observed cold gas.
	
	\subsubsection{Field stability parameter}
	To test other indicators for the thermal stability of the hot gas, we also calculated the Field stability parameter \citep{Field1965} defined as
	
	\begin{equation}
	\Pi_{\mathrm{F}} \equiv \frac{\kappa T}{n_{\mathrm{e}}n_{\mathrm{H}}\Lambda(T)r^2},
	\end{equation}
	
	\noindent where $ \kappa $ is the Spitzer thermal conductivity and $ \Lambda(T) $ is the cooling function. It is as a~measure of the prevalence of the conductive heating rate over the radiative cooling rate on scales close to $ r $. The limiting value for $ \Pi_{\mathrm{F}} $ based on a~sample of 46 brightest cluster galaxies examined in \citet{Voit2008}, below which the thermal conduction is not capable of suppressing radiative cooling, is $ \Pi_{\mathrm{F}} \lesssim 5 $.
	
	As can be seen in the top panel of Figure \ref{fig:fieldviscos}, where the observationally suggested threshold is visualised through the dashed grey line, it is exceeded at most radii in NGC~7049. This result is comparable to the elliptical galaxy NGC~6868, which also shows the presence of a \textit{rotating} disc of cool gas \citep{Werner2014} surrounded by an extended halo of hot  plasma. \citet{Werner2014} propose that rotation might be important for the development of cooling instabilities in these systems.  We caution that there are strong indications that the conductivity of the hot plasma in clusters of galaxies and giant ellipticals is very low (with estimated suppression factors $ \lesssim 10^{-2}$), making conductive heating in these systems potentially irrelevant and the Field criterion invalid \citep[e.g.][]{gaspari2013, DeGrandi2016, Eckert2017}. If the gas is not thermally unstable by the Field criterion, then cold clouds are also generally prone to destruction by evaporation. The survival of a cold/warm phase embedded in the hot plasma also indicates that conduction is suppressed.
	
	\subsubsection{Viscous stability parameter}
\begin{figure}
		\centering
		\includegraphics[width=1.\linewidth]{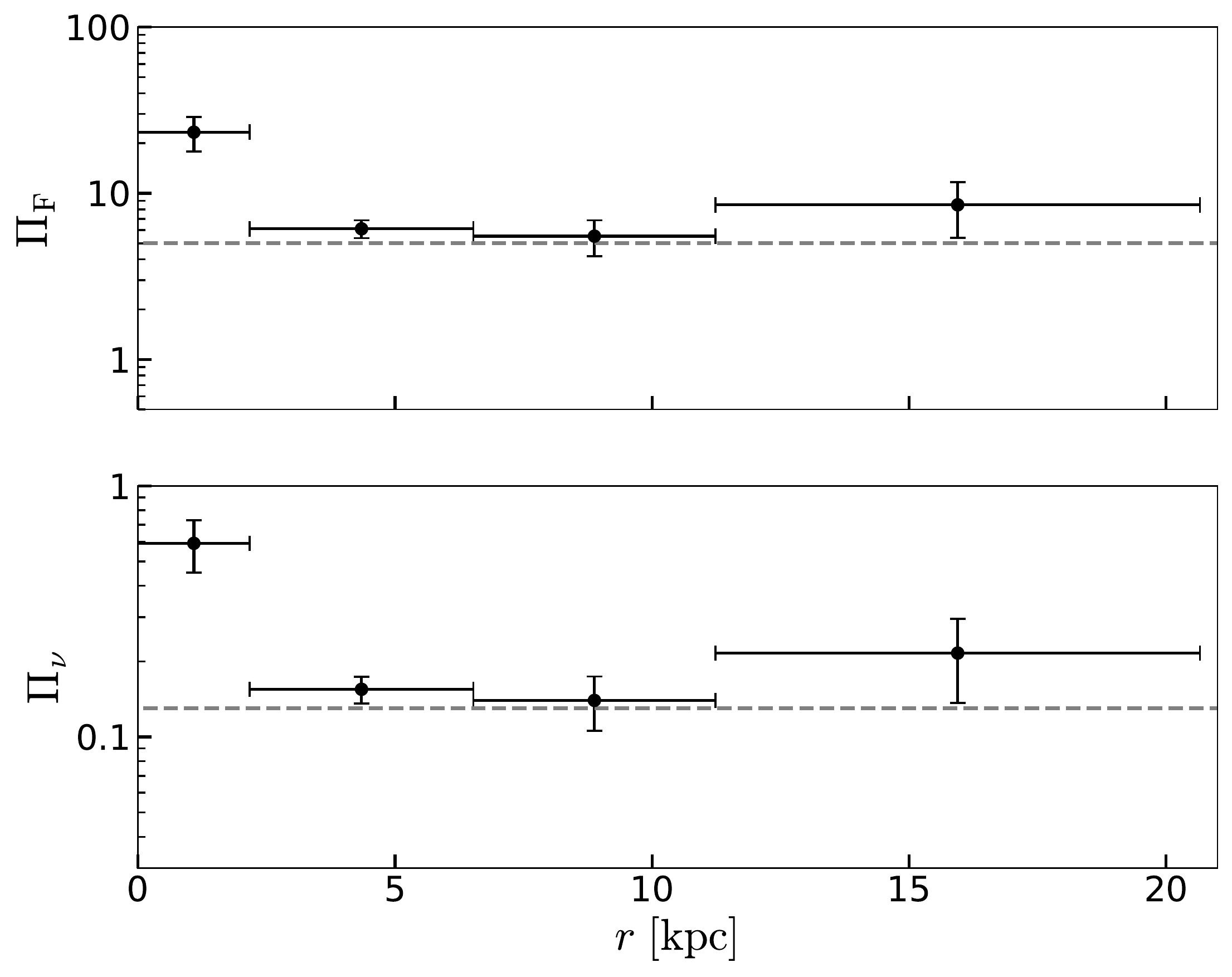}
		\caption{Field stability parameter (top) and viscous stability parameter (bottom) determined from deprojected spectra as a~function of radius. The dashed line in the upper panel shows the observationally suggested threshold \citep[$\Pi_{\mathrm{F}}\lesssim5$,][]{Voit2008}, below which thermal conduction is not capable of suppressing radiative cooling. In the lower panel, a similar threshold is approximately equal to $0.13$. In neither case are the measured stability parameters low enough for us to expect that the hot gas will be thermally unstable.}
		\label{fig:fieldviscos}
	\end{figure}

	Another stability parameter that should be more robust in rotating systems in determining the conditions required for thermal instabilities to develop is the viscous stability parameter, $\Pi_{\mathrm{\nu}}$. This criterion, set by the requirement that the gas can retain the bulk of its angular momentum while it cools, takes into account the viscosity of the cooling medium. It is defined as
	
	\begin{equation}\label{eq:visc}
	\Pi_{\mathrm{\nu}} \equiv \frac{\nu t_{\mathrm{cool}}}{r^2},
	\end{equation}
	
	\noindent where $ \nu $ is the kinematic viscosity of the gas and $ t_{\mathrm{cool}} $ and $ r $ are the cooling time and radius, respectively. It introduces the viscous diffusion length in a~cooling time (square root of the numerator in \ref{eq:visc}) and compares it to $ r $. The parameters $ \Pi_{\mathrm{F}} $ and $ \Pi_{\mathrm{\nu}} $ are not completely independent, as the processes they are based on are both due to Coulomb collisions of either electrons or ions. Their ratio is $ \Pi_{\mathrm{\nu}}/\Pi_{\mathrm{F}} \simeq 0.0253$, so that the previous constraint, $\Pi_{\mathrm{F}}\lesssim5$, corresponds to $\Pi_{\mathrm{\nu}} \lesssim 0.13$. Results from the calculation of the viscosity parameter are presented in the bottom panel of Figure \ref{fig:fieldviscos}, where the dashed grey line symbolises the critical value. We again caution that observations suggest that the viscosity in the intracluster medium is suppressed by at least a factor of ten \citep[e.g.][]{werner2016, Ichinohe2017, Su2017}, which could also affect the applicability of the viscosity criterion. 
	
	\subsubsection{Turbulent Taylor number}
	A key criterion for rotating hot atmospheres has been presented by \citet{Gaspari2015} in the form of the turbulent Taylor number  
	
	\begin{equation}
	\mathrm{Ta_t} \equiv v_{\rm rot}/\sigma_{v}.
	\end{equation}
	
	The rotational velocity and the velocity dispersion were retrieved directly from the $\mathrm{[C\,\textsc{ii}]}$ \textit{Herschel} data cube (assuming that the same dynamical conditions apply in the hot gas). The latter has been corrected for the line broadening contamination due to rotation, $ \sigma_{\rm rot} $ associated with the large extraction beam as follows: We constructed a 3D model of a synthetic disc which had identical inclination, position angle, surface brightness, and velocity gradient. The thickness of the synthetic disc was set to $ 0.5~\rm kpc $, but we note that it had only a negligible effect on the resulting values. From this model, we retrieved $ \sigma_{\rm rot} $ for each pixel of \textit{Herschel} data by doing a luminosity-weighted projection along the line of sight of the velocity variance convolved with the \textit{Herschel} (gaussian) beam (corresponding to FWHM of $12~\rm arcsec$). As expected by simple analytic calculation, this contamination is comparable to roughly $1/3$ of the velocity gradient, which could be used in future samples to quickly remove $ \sigma_{\rm rot} $.
	
\begin{figure}
		\centering
		\includegraphics[width=1.0\linewidth]{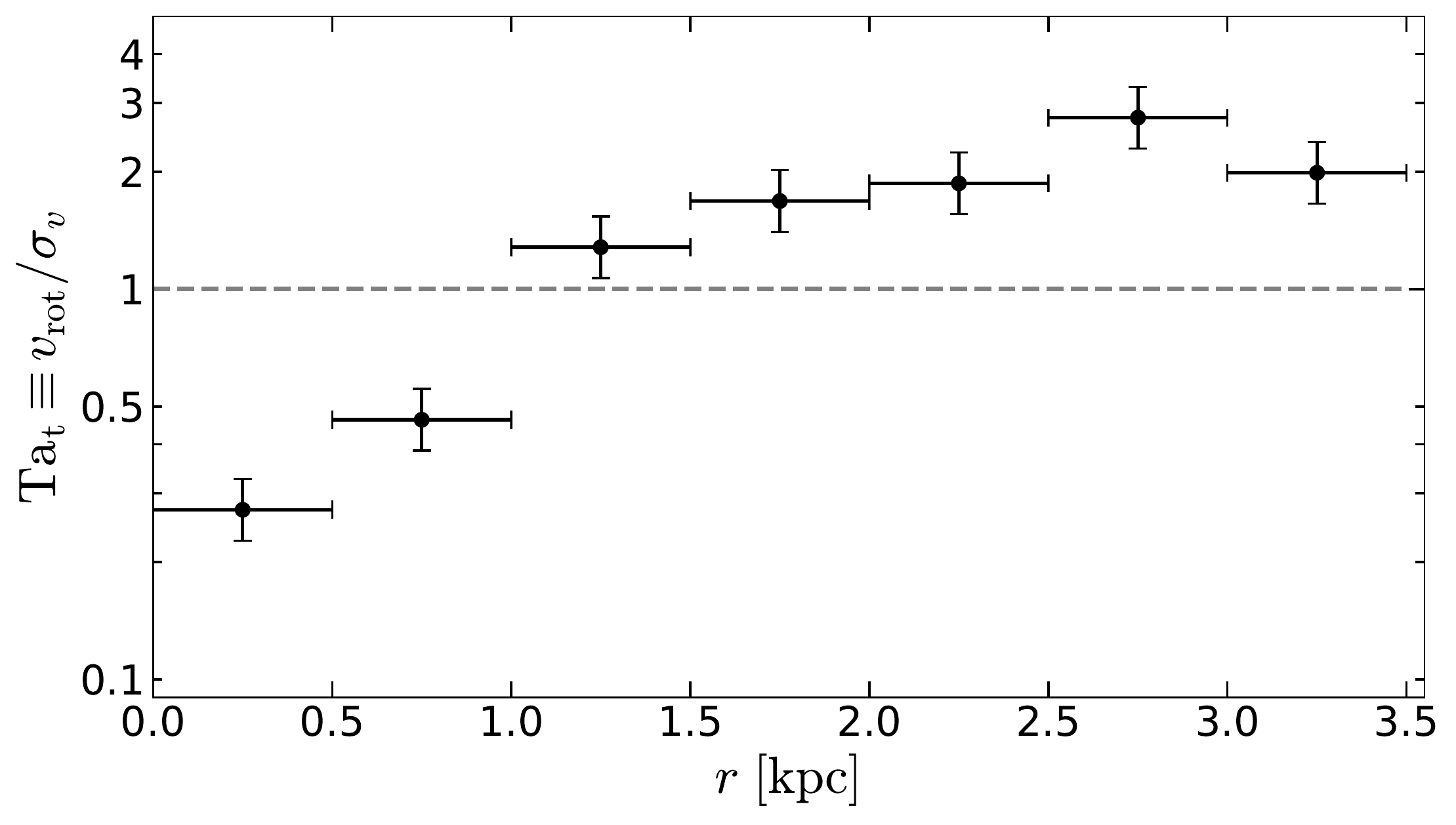}
		\caption{The radial profile of the ratio of rotational velocity and the velocity dispersion for the $\mathrm{[C\,\textsc{ii}]}$-emitting gas. Dashed grey line signifies transition from rotation- to turbulence-dominated dynamics and type of condensation.}
		\label{fig:Ta_t}
	\end{figure}

	The radial profile of $\rm Ta_t$ is shown in Figure \ref{fig:Ta_t}. It is evident that NGC~7049 has $\rm Ta_t >1 $ over most of the volume, while only the inner $1~\mathrm{kpc}$ shows a ratio $<1$. This implies that the gas condensing from the hot halo will follow helical paths, settling onto the equatorial plane and thus forming a kpc-scale multiphase disc. In the nuclear region, turbulence becomes relatively more prominent compared with rotation, thus it may trigger a phase of CCA rain and boosted AGN feeding in the near future.

	\subsubsection{C-ratio}
	A criterion closely related to presence of turbulent motions in the gas and the related condensation cascade is the so-called $C$-ratio \citep{Gaspari2018}
	
	\begin{equation}
	C \equiv \frac{t_{\rm cool}}{t_{\rm eddy}},
	\end{equation}
	
	\noindent where the eddy turnover time is
	
	\begin{equation}
	t_{\rm eddy} = 2\pi \frac{r^{2/3} L^{1/3}}{\sigma_{v,L}},
	\end{equation}
	
	\noindent with $\sigma_{v,L} $ the velocity dispersion at the injection scale length $L$, which can be estimated by the diameter of the cold/warm gas emission ($L \sim 7~\rm kpc $). This can be obtained from the observed line-of-sight velocity dispersion profile (corrected for the line broadening due to rotation), such as $ \sigma_{v,L} = \sqrt{3} \sigma_{v,\rm los}(L)$. This time-scale is related to the generation of density fluctuations driven by the turbulent eddies in a stratified environment and should be comparable to the cooling time in order for non-linear condensation to develop, that is $C \approx 1$. We note that given the dominance of rotation (see $\rm Ta_t$ ratio above), the condensed gas does not form filaments, but it is forced to settle onto the equatorial region augmenting the extended multiphase disc (which could still have some intrinsic turbulence). 
	
	\begin{figure}
		\centering
		\includegraphics[width=1.0\linewidth]{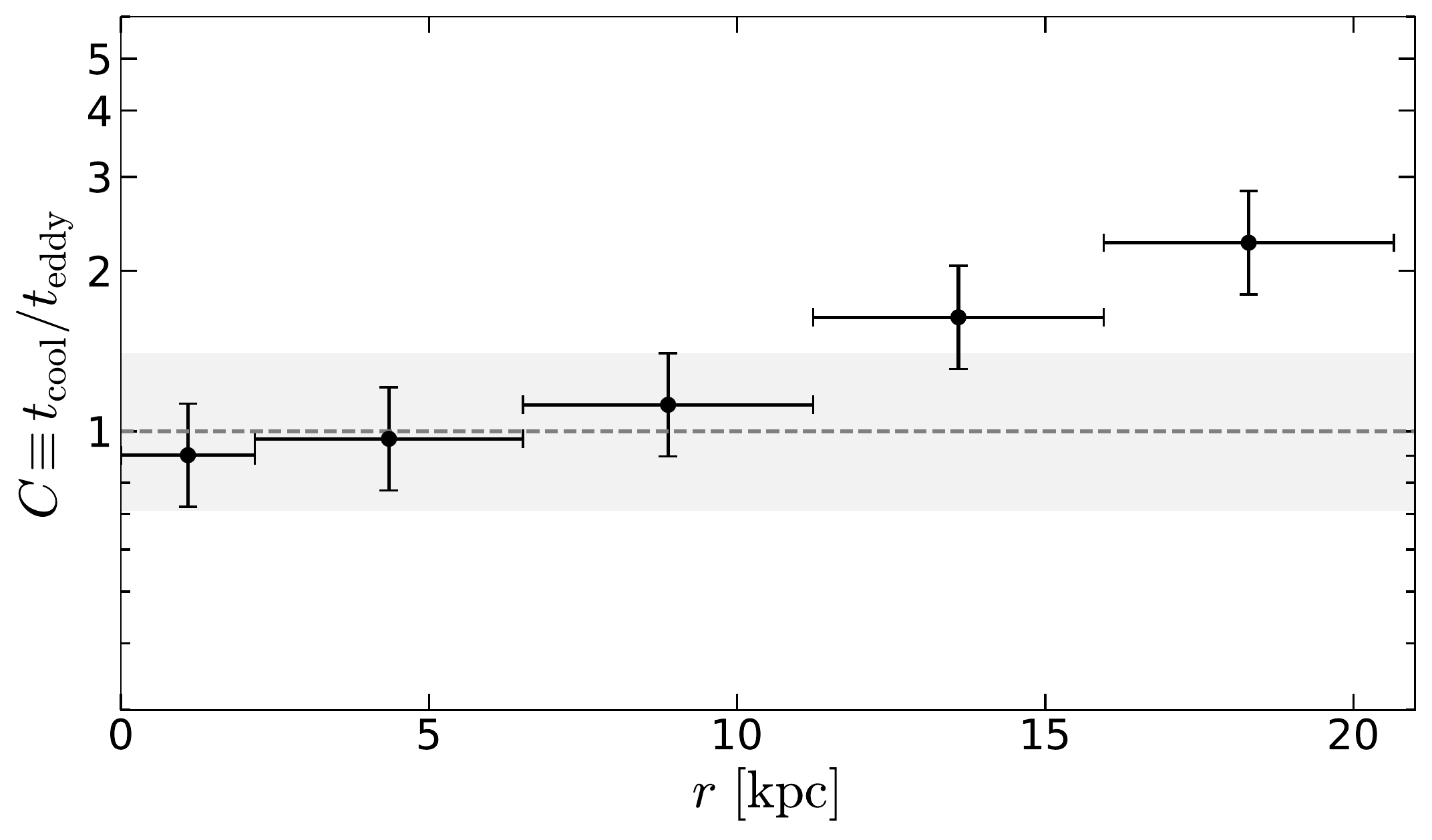}
		\caption{The radial distribution of the cooling time over eddy turnover time for NGC 7049. Grey region represents the 1-$\sigma$ confidence region (retrieved from hydrodynamical simulations; \citealt{Gaspari2018}) for significant multiphase condensation to develop.}
		\label{fig:C_ratio}
	\end{figure}

	\section{Discussion}\label{sec:discussion}
	
	\subsection{Shape of the X-ray halo}
	The image of the X-ray-emitting gas is remarkably smooth and shows no detectable traces of recent violent processes. Based on image analysis, we find that the X-ray emission has an ellipticity of $ \epsilon_{\rm X} = 0.126\pm0.004 $, which is by a factor of two lower than the ellipticity of the co-oriented stellar component $  \epsilon_{\star} = 0.25 \pm 0.02 $. We note that the observed ellipticity might be underestimated due to the inclination of the system. However, the results of azimuthally resolved spectral analysis  show no systemic difference in density in the plane of rotation compared to the perpendicular direction. Numerical simulations of \cite{Brighenti2009} suggest that the small ellipticity can be explained by strong gas motions. The presence of turbulence would lead to more spherically symmetric atmosphere, as observed here. Our constrained subsonic turbulence has a mixing time ($t_{\rm mix} \simeq t_{\rm eddy} $) less than a Gyr within $ r \lesssim 10~\mathrm{kpc} $, implying that turbulent mixing is indeed impactful over the cosmological evolution of NGC~7049. In passing, we note that subsonic turbulence is a typical feature in giant ellipticals \citep{werner2009,deplaa2012,ogorzalek2017}.
	
	\citet{Hanlan2000} studied a sample of 6 nearby objects with various rotational velocities and found that fast rotating elliptical galaxies have a smaller  ellipticity in the X-rays than in the optical band. On the other hand, the lenticular galaxy NGC~6868 shows flattened  X-ray isophotes \citep{Machacek2010}, but it is currently also undergoing a~merger.
	
	Compared to the elliptical galaxies in the sample of \citet{OSullivan2003}, the X-ray surface brightness profile of NGC~7049 is shallow, and it is comparable to the shallower X-ray brightness profiles of spirals in the sample of \citet{Li2017}.
	
	\subsection{Spectral properties}
	From the global spectrum, we estimated the metallicity of the gas to be $ Z = 0.7^{+0.2}_{-0.1}~Z_{\odot}$ which is comparable to other early-type galaxies. However, this result might be affected by our assumption of Solar relative abundances (abundances of various metals relative to Fe) in the gas and a constant metallicity over the X-ray halo, which is usually not observed \citep[e.g.][]{Tozuka2008}. 
	
	A biased metallicity would also affect our measured normalisations and densities. A factor of two difference between the real and measured metallicity would result in a $25$ per cent bias in the density. The inferred pressure would be affected by the same factor and the entropy would be biased by $17$ per cent. Gradients in metallicity would alter the measured slopes of the thermodynamic quantities by less than $10$ per cent \citep{Werner2012}.
	
	The emission-weighted temperature and the temperature profile of NGC~7049 are within the range observed for a sample of 53 elliptical galaxies by \citet{Fukazawa2006}. The authors proposed a density criterion $ n_{\mathrm{e}}(r=10~\mathrm{kpc}) < 3\times 10^{-3}~\mathrm{cm}^{-3}$, where NGC~7049 fits among the low-density objects, which can display positive, negative, or variable temperature profiles. The global emission-weighted temperature is also comparable to those measured for a sample of rotating galaxies observed with the \textit{Chandra X-ray Observatory} \citep{Posacki2013}. In \citet{Diehl2008}, the projected outer temperature profiles of ellipticals are found to be set by their environment, so that the negative gradients appear to be linked to low-densities of the surrounding medium, while positive gradients are found in galaxies in groups and clusters. The projected temperature profile of NGC~7049 would thus indicate a relatively low-density environment. The negative gradient is also a typical sign of compressional heating which dominates the lower potentials of low-mass groups or isolated
	early-type galaxies \citep[e.g.][]{Gaspari2012b}. 
	
\begin{figure}
		\centering
		\includegraphics[width=1\linewidth]{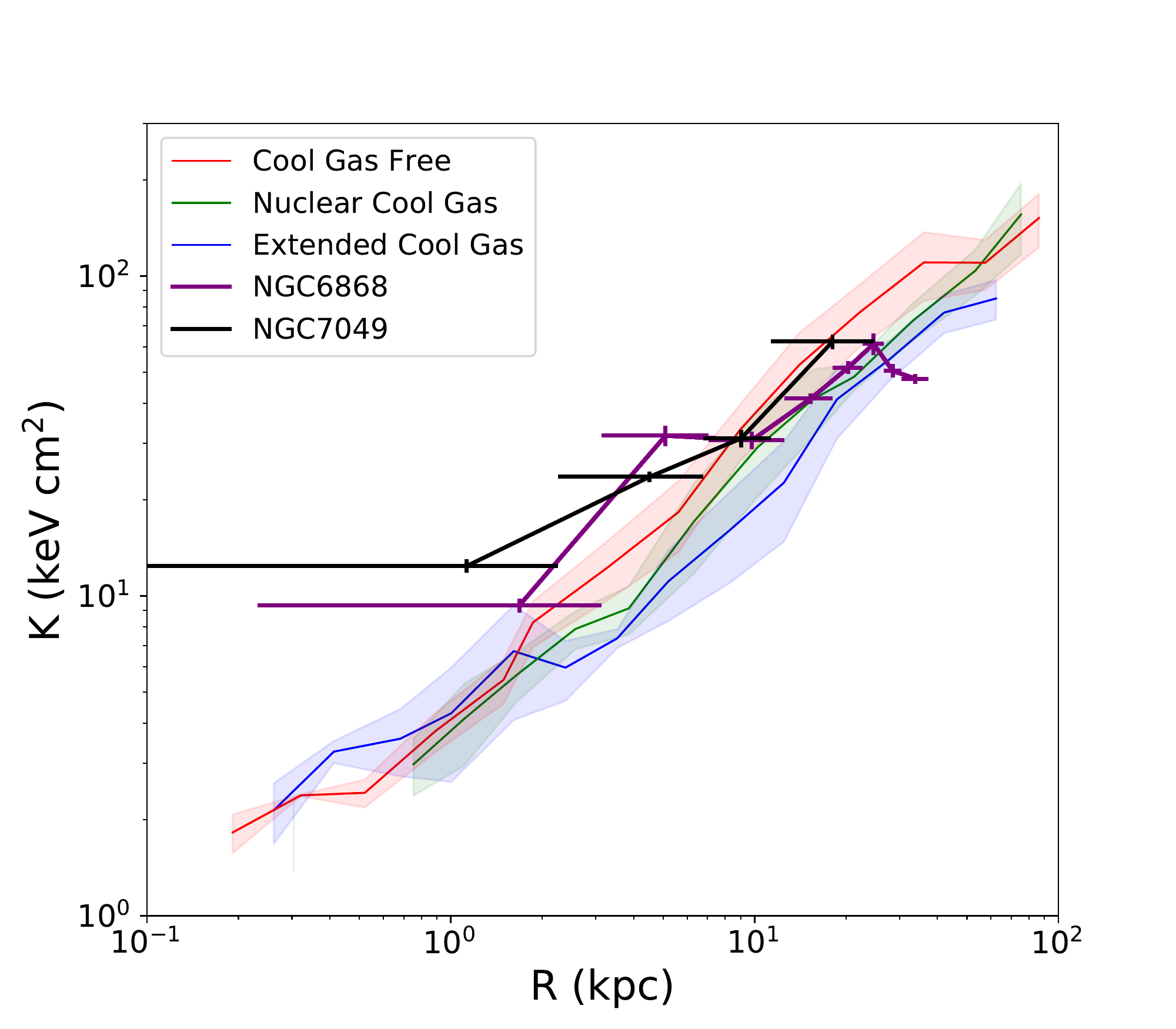}
		\caption{The median entropy profile for a sample of 49 galaxies from \citet{lakhchaura2018} with the entropy profiles of NGC~7049 and another galaxy with a fast rotating disc of cool gas, NGC~6868, overplotted.  The red, green and blue solid lines show median profiles for the cool gas free, nuclear cool gas and extended cool gas systems, respectively and the shaded regions show the median absolute deviation (MAD) spreads about the medians.}
		\label{fig:KL_entropy}
	\end{figure}

	We also compared our entropy profile with the galaxies in the work of \citet{lakhchaura2018}. They studied the thermodynamic properties of the hot atmospheres of galaxies with extended and nuclear $\mathrm{H}\,\alpha $+$[\mathrm{N\,\textsc{ii}}]$ emission and without any detectable emission line nebulae. In Figure \ref{fig:KL_entropy}, we overplotted the entropy profile of NGC~7049 and NGC~6868, which  also harbours a rotating disc of cool gas \citep[see][]{Werner2014}, on the set of median entropy profiles for a sample of 49 galaxies from \citet{lakhchaura2018}. Their sample indicates that the cool gas free systems have, on average, systemically higher entropies than the systems with extended filamentary $\mathrm{H}\,\alpha $+$[\mathrm{N\, \textsc{ii}}]$ nebulae. 
	At small radii, NGC~7049 and NGC~6868, which both harbour extended disc-like (not filamentary) $\mathrm{H}\,\alpha $+$[\mathrm{N\, \textsc{ii}}]$ emission, have profiles comparable to the higher entropy, cool gas poor, non-rotating galaxies. The entropy profile of NGC~7049 is flatter than that of NGC~6868 and its innermost value is higher than that of any other system in the \citet{lakhchaura2018} sample. The high central entropy and the negative temperature profile indicate that a centrally positioned heating source is present in the galaxy. When the central entropy of the gas increases, the pressure of the surrounding medium can lead to an increase of the gas temperature and subsequently its observed radially decreasing trend. It also suggests that the hot gas is not convectively stable and the X-ray atmosphere could be overheated and expanding.

	\subsection{Cooling process}
	
	An estimate of the amount of cold gas in the plane of the rotating disc in NGC~7049 can be made from $\mathrm{[C\,\textsc{ii}]\lambda157~\umu m} $ line. In normal galaxies and Galactic molecular clouds, the $\mathrm{[C\,\textsc{ii}]\lambda157~\umu m} $ emission is about $1500 \times$ stronger than the $\mathrm{CO(J = 1 \rightarrow 0)} $ line, which can then be used in standard CO luminosity to $ \rm H_2 $ mass conversion. The integrated $\mathrm{[C\,\textsc{ii}]} $ line intensity of $ 2.24 \times 10^{-13}~\rm erg~s^{-1}~cm^{-2}$ \citep{Werner2014} would then give a molecular gas mass of about $ 5 \times 10^{7}~\mathrm{M_{\odot}} $. This value should serve only as a very rough approximation, as the temperature and the density of the molecular gas may differ significantly from the numbers assumed in the calculation.
	
	In the regions associated with the cold-gas disc, we have detected the presence of multi-temperature X-ray gas, indicating ongoing cooling and thus suggesting that the X-ray gas could be a possible origin of the cold phase. However, the most commonly used criteria for cooling instability are not fulfilled in this system. 
	
	The TI-ratio $\lesssim10$ criterion \citep[e.g.][]{Sharma2012} assumes that the gas is mainly supported by the hydrodynamic buoyant force against gravity. If condensation develops, the gas entropy locally decreases and the condensed gas starts moving inwards into the lower entropy medium. If the infall speed of the perturbed gas is fast enough, it can reach a location where the heating rate is sufficient to prevent further cooling before cooling runs away.
	
	This is where the rotational support might be crucial. The time for the cooling clump of gas with non-zero angular momentum to move inwards is longer than in the absence of rotation allowing the gas to cool regardless of the restoring buoyant force (see Sect. 3 in \citealt{Gaspari2017}). The criterion of $ t_{\mathrm{cool}}/t_{\mathrm{ff}}\lesssim 10 $ would thus not have to be strictly followed and a coherent extended condensation leading to a multiphase disc would develop even for $ t_{\mathrm{cool}}/t_{\mathrm{ff}}> 10$. The actual threshold value would then also depend on other processes that have influence on mixing of the cooling gas with the surrounding environment and on the dissipation of angular momentum, such as viscosity and turbulence.
	
	The $C$-ratio, in which rough equality of the gyration timescale of individual turbulent eddies and the cooling timescale is considered to be crucial for the development of non-linear thermal instability and onset of cooling \citep{Gaspari2018}, is consistent with the observed X-ray gas properties. Along with the $ \rm Ta_t $ radial profile, which suggests relative dominance of rotation at distances exceeding $ r \sim 1 ~ \rm kpc $, the cooling gas should be moving towards the equatorial plane following conical helical paths.
	Considering currently suppressed accretion onto the central SMBH (see Sect. \ref{sec:radial}), the declining $ \rm Ta_t $ towards the centre of the galaxy may forecast a new generation of central condensation and boosted AGN activity.
	
	We also consider the alternative, that the cool dusty gas could have originated through stellar mass-loss or that it could have been brought by a merger event. The evolved stellar population of the galaxy is expected to provide $\sim1~\mathrm{M}_{\odot}~\mathrm{yr}^{-1}$ of gas per stellar mass of $10^{11}~\mathrm{M}_\odot$ \citep{canning2013}. Most of this stellar mass loss material is expected to thermalise to the virial temperature of the system \citep[see e.g.][]{parriott2008,bregman2009}, however a significant fraction might be also contributing to the cool dusty gas mass budget.  
	Observations of NGC~7049 in $\mathrm{H}\,\alpha  $ and $[\mathrm{N\,\textsc{ii}}]$ carried out by the Very Large Telescope and analysed in \citet{Coccato2007} reveal the presence of an inner disc of ionized gas orthogonal to the main sense of rotation that is a~few arcsec wide. The fact that the gas in the inner polar disc is geometrically decoupled from the main one, indicates that it cannot be attributed to a~single gaseous component of a~strongly warped disc. As \cite{Coccato2007} state, even though such features are not rare among lenticular galaxies, it cannot be ruled out that it has been created by accretion of matter from an infalling galaxy. A past merger event could have contributed to the creation of the cold disc, but an increase of central entropy and negative temperature gradient should not be regarded as corollary of a merger, as there is no sign of a large-scale perturbation in the hot gas. Such a merger scenario would, however, not exclude the possibility of ongoing cooling and it would not explain the presence of multi-temperature X-ray gas in the plane of the disc. 
	
	Regardless of its origin, the presence of the cool gaseous disc suggests that the AGN avoids destroying it. AGN driven feedback in the form of collimated outflows or jets propagating mainly perpendicular to the disc without dissipating in its vicinity would be capable of preserving it, regardless of the fact that much more heat would need to be injected to the cold gas in order to increase its temperature than to the less dense hot regions.  Rotational support of the gas would also alter the accretion rate and feeding of the AGN, leading to further development of the disc undisturbed by AGN winds and jets. 
	
	\section{Conclusions}\label{sec:conclusions}
	Our analysis of the {\it XMM-Newton} observation of the massive fast-rotating lenticular galaxy NGC~7049 has led to the following results. 
	\begin{itemize}
		\item The X-ray spectral modelling properties are: emission-weighted temperature $ k_{\mathrm{B}}T = 0.43^{+0.02}_{-0.01}~\mathrm{keV}$; emission-weighted metallicity: $ Z = 0.7^{+0.2}_{-0.1}~Z_{\odot}$; ellipticity $ \epsilon_{\rm X} = (0.126\pm0.004) $; and central density $ n(0) = (0.031 \pm 0.001)~ \rm cm^{-3}$.
		\item The hot gas has an unusually high central entropy and a temperature peak.
		\item While the hot gas in the rotational plane of the cool dusty disc has a multi-temperature structure, the thermal structure along the
		rotation axis is single-phase. The observed azimuthal difference in the temperature structure indicates that cooling is more efficient in the equatorial plane, where the rotational support of the hot gas may be able to alter the condensation, regardless of the $  t_{\mathrm{cool}}/t_{\mathrm{ff}} $ criterion, which is here relatively high ($\sim40$). 
		\item We analysed other criteria for multiphase gas formation and evolution, finding $C$-ratio $ \approx 1 $, which implies significant condensation, and $ \rm Ta_t > 1$, which indicates such a condensation occurs onto non-radial orbits forming a disc (instead of filaments). This is in agreement with hydrodynamical simulations of massive rotating galaxies predicting a similarly extended multiphase disc \citep[e.g.][]{Gaspari2017}.  
	\end{itemize}
	
	\section*{Acknowledgements}
	
	This work was supported by the Lend\"ulet LP2016-11 grant awarded by the Hungarian Academy of Sciences. M. G. and R.E.A.C.  are supported by NASA through Einstein Postdoctoral Fellowship Award Number PF5-160137 and PF5-160134, respectively,  issued by the Chandra X-ray Observatory Center, which is operated by the SAO for and on behalf of NASA under contract NAS8-03060. A. S. is supported by the Women In Science Excel (WISE) programme of the Netherlands Organisation for Scientific Research (NWO).
	Part of this work was carried out during the Undegraduate Summer Research Program 2018 in the Department of Astrophysical Sciences at Princeton University.
	Based on observations obtained with XMM-Newton, an ESA science mission with instruments and contributions directly funded by ESA Member States and NASA.
	This research has made use of data obtained from the Chandra Data Archive, and software provided by the Chandra X-ray Center (CXC) in the application packages CIAO and Sherpa.
	We thank the anonymous referee for the constructive feedback which helped to improve the manuscript.
	
	
	
	
	\bibliographystyle{mnras}
	\bibliography{ms}
	
	
	
	%
	%
	

	\label{lastpage}
\end{document}